\documentclass[aps,preprint]{revtex4}
\usepackage{graphicx,amsmath}

\draft
\begin{document}
\author{Min He$^{a}$, Fei Hu$^{a}$, Wei-Min Sun$^{a,b}$, and Hong-Shi Zong$^{a,b}$}
\address{$^{a}$ Department of Physics, Nanjing University, Nanjing 210093, P. R. China}
\address{$^{b}$ Joint Center for Particle, Nuclear Physics and Cosmology, Nanjing 210093, China}
\title{Crossover from a continuum study of chiral susceptibility}
\begin{abstract}
We derive a model-independent integral formula for chiral
susceptibility and attempt to present a continuum model study of it
within the framework of Dyson-Schwinger Equations. An appropriate
regularization is implemented to remove the temperature-independent
quadratic divergence inherent in this quantity. While it
demonstrates a second-order phase transition characteristic in the
chiral limit, the result obtained supports a crossover at physical
current quark masses, which is in good agreement with recent lattice
studies.
\bigskip


\bigskip
 E-mail: zonghs@chenwang.nju.edu.cn.

\bigskip


\end{abstract}
\maketitle
\section{Introduction}
Quantum Chromodynamics (QCD) underlies our current understanding of
strong interaction and explains why and how the fundamental
constituent, namely quarks and gluons are bound into nucleons,
pions, etc. The QCD vacuum is rather complex and intriguing: while
it features the overwhelming spontaneous chiral symmetry breaking
and color confinement, it is believed to be heated and undergo a
transition at high enough temperature \cite{a1,a2,a3,a4}. As a
result, quarks and gluons are no longer confined inside hadrons and
a new state of strongly interacting matter, the so-called
quark-gluon plasma (QGP) is expected to form. Meanwhile, the chiral
symmetry gets restored. It is a goal to detect such a new state of
matter for the on-going heavy-ion collision experiments at the BNL
Relativistic Heavy-Ion Collider (RHIC) and the near future Large
Hadron Collider (LHC). It is widely speculated that this QCD-related
transition may also be of direct relevance for the evolution of the
early Universe \cite{a1,a5}

Although lattice QCD has established the existence of such a thermal
transition, the nature of this transition (its order, or, whether is
is a real phase transition) has been a long time debate \cite{a1}. It
is not until recently that Y. Aoki and his collaborators, by
exploiting quite demanding and powerful lattice simulations, assert
unambiguously that QCD evolves smoothly with temperature; that is,
there is no thermodynamic phase transition for three flavors (u, d,
s-quark) with physical current masses, but instead a smooth
crossover \cite{a2,a6}. Such a conclusion rules out some scenarios
and speculations over the evolution of the early Universe \cite{a1}.
In fact, the nature of the QCD thermal transition would lay an
imprint on the Universe's evolution and thus affect our
understanding of the latter. For example, the previously prevailing
strong first-order transition scenario predicts inhomogeneous
nucleosynthesis and later formation of cold dark matter
clumps \cite{a5,a7}. Nowadays, people tend to believe that the nature
of QCD thermal transition depends on the quark content and current
quark masses \cite{a8,a9}. In the chiral limit,
the chiral phase transition is likely to be of second order for two
flavors and first order for three flavors. For infinite or very
large current quark masses, the transition turns out to be a
first-order deconfinement transition. For intermediate current masses,
one may end up with a crossover, instead of a real phase transition,
meaning that thermodynamic quantities would show a rapid change, as
opposed to a jump, in a narrow temperature range.

As is well known, first principle lattice simulations provide the
prior choice to address the thermal QCD transition problem, because
of its intrinsic nonperturbative nature. In this regard, recently Y.
Aoki and collaborators \cite{a6} achieved a large scale computing and
identified the nature of three-flavor QCD finite temperature
transition as a smooth crossover at physical quark masses. On the
other hand, other than lattice simulations, continuum
nonperturbative model study can be a complementary approach to
thermal QCD transition. Following Y. Aoki et al. in
Refs. \cite{a2,a6}, we present a continuum model study of the interesting
chiral susceptibility in the present work, from whose behavior as
temperature varies we identify the nature of two-flavor QCD thermal
transition in the case of physical current quark masses as well as
in the chiral limit.

We first derive a model-independent analytical formula, which
expresses chiral susceptibility as a integral in terms of
dressed propagators and dressed scalar vertex. The quadratic
divergence inherent in this susceptibility is then demonstrated and
an appropriate renormalization procedure is specified. In the
section that follows, this result is then illustrated by a model
calculation based on an effective interaction in the framework of rainbow Dyson-Schwinger equation (DSE)
and ladder Bethe-Salpeter equation (BSE) approach. The notorious
quadratic ultraviolet divergence is appropriately subtracted by a
delicate numerical procedure. The results obtained are carefully
analyzed and a crossover and a second-order phase transition are
respectively identified for the cases of physical and zero current
quark masses. The final section is devoted to the conclusion and
discussion. Throughout this paper, we work with Euclidean space
metric: $\{\gamma_\mu,\gamma_\nu\}=2\delta_{\mu\nu}$.

\section{DERIVATION AND RENORMALIZATION}
\subsection{Derivation of an integral formula for chiral susceptibility}
Chiral susceptibility has interested quite a few lattice QCD
practitioners and attracted much investigation \cite{a3,a6,a8,a9,a10,a11,a12,a13,a14,a15,a16,a17,a20}.
It is defined as
\begin{eqnarray}
\chi=\frac{T}{V}\frac{\partial^2}{\partial m^2}\log Z
=(-)\frac{\partial}{\partial m}<\overline{\psi}\psi>,
\end{eqnarray}
where $Z$ denotes the QCD partition function at finite temperature
and vanishing chemical potential.
\begin{equation}
Z=\int\mathcal{D}\overline{\psi}\mathcal{D}\psi\mathcal{D}A
exp\{-\int^\beta_0d\tau\int
d^3x[\overline{\psi}_f(\gamma\cdot\partial+m_f+ig\frac{\lambda^a}{2}\gamma\cdot
A^a)\psi_f+\frac{1}{4}F^a_{\mu\nu} F^a_{\mu\nu}]\}.
\end{equation}
Here, $\overline{\psi}$, $\psi$ and $A^a_\mu$ are subject to
anti-periodic and periodic conditions, respectively; and we leave
the gauge fixing term and the ghost field term to be understood.
Note that chiral susceptibility is defined for two light flavors and
hence $m$ in Eq. (1) is supposed to be the degenerate current mass of
u, d quarks. The second line of that equation also indicates that
the chiral susceptibility measures the response of chiral
condensate (the order parameter) to a small perturbation of the
parameter responsible for explicit breaking of chiral symmetry (the
current quark mass).

Formally, using quark propagator we can always write the quark
condensate as
\begin{equation}
<\overline{\psi}\psi>=(-)N_cN_f\int\frac{d^4p}{(2\pi)^4}tr_\gamma
G(p,m),
\end{equation}
where $tr_\gamma$ denotes trace over Dirac indices of the quark
propagator, $N_c=3$ is the color factor and $N_f=2$ denotes two
degenerate light flavors. Substituting Eq. (3) into Eq. (1) and
adopting the identity
\begin{equation}
\frac{\partial G(p,m)}{\partial m}=-G(p,m)\frac{\partial
 G^{-1}(p,m)}{\partial m}G(p,m),
\end{equation}
we have
\begin{equation}
\chi=(-)N_cN_f\int\frac{d^4p}{(2\pi)^4}tr_\gamma[G(p,m)\frac{\partial
G^{-1}(p,m)}{\partial m}G(p,m)].
\end{equation}
We consider the mass $m$ as a constant background scalar field
coupled to the quark fields by the term $m\overline{\psi}\psi$. Then
$G(p,m)$ is the dressed quark propagator in the presence of such a
background field and the derivative of its inverse with respect to
the current quark mass yields the so-called one-particle-irreducible (1PI) dressed scalar vertex (a three-point correlation function)
\begin{equation}
\frac{\partial G^{-1}(p,m)}{\partial m}=\Gamma(p,0;m),
\end{equation}
where $p$ is the relative momentum. This relation may be called
``scalar Ward identity''. Note that the total momentum of the dressed
scalar vertex vanishes because the background scalar field $m$ is a
constant that takes no momentum. Substituting Eq. (6) into Eq. (5) gives
\begin{equation}
\chi=(-)N_cN_f\int\frac{d^4p}{(2\pi)^4}tr_{\gamma}[G(p,m)\Gamma(p,0;m)G(p,m)].
\end{equation}
This is the integral formula for chiral susceptibility at zero
temperature. Diagrammatically, the right-hand-side of Eq. (7) expresses the Feynman diagram for the color-singlet scalar vacuum polarization at zero total momentum. This clearly shows that chiral susceptibility at zero temperature equals the color-singlet scalar vacuum polarization at zero total momentum.

According to finite-temperature field theory (see, e.g.,
\cite{a18}), the corresponding finite temperature version is
obtained by replacing the integration over the fourth component of
momentum with summation over fermion Matsubara frequencies:
\begin{equation}
\chi(T)=(-)N_cN_fT\sum\limits_{n=-\infty}^{+\infty}\int\frac{d^3p}{(2\pi)^3}tr_{\gamma}[G(p_n,m)\Gamma(p_n,0;m)G(p_n,m)],
\end{equation}
where $p_n=(\vec{p},\omega_n)$ with Matsubara frequencies
$\omega_n=(2n+1)\pi T$. Here it should be noted that in the chiral
limit the above expression for the chiral susceptibility reduces to
the corresponding expression (Eq. (7)) in Ref. \cite{a19}.

So we have derived a model-independent integral formula which
expresses the chiral susceptibility in terms of the dressed quark
propagator and the dressed scalar vertex, both of the latter being
basic quantities in quantum field theory (QFT). The DSE-BSE
approach provides a desirable framework to calculate these
quantities non-perturbatively and hence the chiral susceptibility.

\subsection{Quadratic divergence and renormalization}
There resides an additive quadratic ultraviolet divergence in the
chiral susceptibility, which can manifest itself even in the chiral
limit \cite{a2,a6}. To see this unambiguously, we calculate the
chiral susceptibility in the case of free quark gas, because all
quantities tend to the free case counterparts in the ultraviolet
limit (asymptotic freedom) and all divergences originate from the
ultraviolet limit. In this case, the dressed scalar vertex reduces
to the bare one, i.e.
\begin{equation}
\Gamma(p_n,0)\rightarrow\mathbf{1},
\end{equation}
and the free quark propagator in the chiral limit reads
\begin{equation}
G^{free}(p_n)=\frac{1}{i\gamma \cdot p_n}.
\end{equation}
Substituting Eqs. (9) and (10) into Eq. (8), one gets
\begin{equation}
\chi^{free}(T)=4N_cN_fT\sum\limits_{n=-\infty}^{+\infty}\int\frac{d^3p}{(2\pi)^3}\frac{1}{\omega_n^2+\vec{p}^2}.
\end {equation}
With the aid of the identity
\begin{equation}
T\sum\limits_{n=-\infty}^{+\infty}\frac{x}{\omega_n^2+x^2}=\frac{1}{2}(1-2n_F(x)),
\end{equation}
where $n_F(x)=\frac{1}{e^{x/T}+1}$ is the Fermi-Dirac statistics
function, we have
\begin{equation}
\chi^{free}(T)=2N_cN_f\int\frac{d^3p}{(2\pi)^3}\frac{1}{|\vec{p}|}(1-2n_F(|\vec{p}|)).
\end{equation}
The second part of the integrand can be integrated out and gives a
purely temperature-dependent finite result: $-\frac{N_cN_f}{6}T^2$.
On the contrary, the first part yields an additive quadratic
divergence. This is in striking contrast to the free quark-number
susceptibility, which is a definitely finite quantity proportional
to $T^2$ and therefore vanishes at zero temperature \cite{a19}.

Actually, there always contains an inherent additive quadratic
divergence in chiral susceptibility. One dose not really have to
address the issue of the regularization of this quadratic
divergence, when one works with a hard cut-off in the numerical
integration (e.g., lattice simulations), and is concerned only with
the temperature-dependent behavior of the susceptibility, not so
much on its absolute value \cite{a21}. However, a correct
renormalization procedure is important and indispensable if one
wants to get correct physics in the continuum limit \cite{a2}. In
Ref. \cite{a19}, we identified the disconnected part of the chiral
susceptibility and argued that it is of major relevance and free
from quadratic ultraviolet divergence. However, we were unable to
derive the disconnected chiral susceptibility directly from a
first-principle approach such as the functional integral method.
Looked at from this point, the identification of the disconnected
chiral susceptibility is not yet absolutely ``waterproof'' and solid
enough. So in this paper we follow the renormalization procedure in
Ref. \cite{a2}, which is to subtract the zero temperature
susceptibility and study the difference between $T\neq0$ and $T=0$.
That is to say, we define the renormalized chiral susceptibility by
\begin{equation}
\chi_R(T)=\chi(T)-\chi
\end{equation}
where $\chi(T)$ and $\chi$ are provided by Eq. (8) and Eq. (7),
respectively. The justification for this renormalization is
two-fold: firstly, the quadratic divergence is additive and
temperature-independent and hence can be really removed by
subtracting the zero-temperature susceptibility; secondly, this
subtraction will never affect the genuine temperature effects of
chiral susceptibility that we are interested in.

Having specified the renormalization procedure, in the next section,
we will focus on the model numerical calculation of the renormalized
chiral susceptibility $\chi_R(T)$ within the framework of DSE-BSE.

\section{MODEL NUMERICAL CALCULATION}
Dyson-Schwinger equations provide a non-perturbative, continuum
approach for the exploration of strong interaction
physics \cite{a21}. Derived from QCD's Euclidean space generating
functional, they constitute a tower of enumerable coupled integral
equations whose solutions are the n-point Green's functions. A
consistent truncation scheme that makes the DSE calculations
tractable is the so-called rainbow-ladder approximation. Such a
rainbow-DSE and ladder-BSE approach has been used extensively in the
investigation of spontaneously chiral symmetry breaking and
confinement \cite{a22} and also found successful applications in
calculating light pseudo-scalar and vector meson
observables \cite{a23,a24} as well as in describing strong
interaction at finite temperature and/or density \cite{a25}. The
subsequent model calculation of chiral susceptibility will also be
conducted systematically and consistently in the rainbow-DSE and
ladder-BSE framework with an effective gluon propagator.

\subsection{The zero-temperature chiral susceptibility}
We first calculate the zero-temperature chiral susceptibility with
Eq. (7). This involves the calculation of dressed quark propagator
and dressed scalar vertex. The former is solved from the rainbow gap
equation
\begin{equation}
G^{-1}(p,m)=i\gamma \cdot
p+m+\frac{4}{3}\int\frac{d^4q}{(2\pi)^4}g^2D_{\mu\nu}^{eff}(p-q)\gamma_{\mu}G(q,m)\gamma_{\nu}
\end{equation}
and the latter satisfies the ladder inhomogeneous BSE (the total
momentum of the dressed scalar vertex has been set equal to zero)
\begin{equation}
\Gamma(p,0;m)=\textbf{1}-\frac{4}{3}\int\frac{d^4q}{(2\pi)^4}g^2D_{\mu\nu}^{eff}(p-q)\gamma_{\mu}G(q,m)\Gamma(q,0;m)G(q,m)\gamma_{\nu},
\end{equation}
where $D_{\mu\nu}^{eff}(p-q)$ is the effective gluon propagator,
which is usually a phenomenological input in practice (it should
also be noted that the analytical structure of the gluon propagator
has been explored from numerical solutions of coupled DSEs of
quarks, gluons and ghosts and from fits to lattice data in recent
literatures, see Ref. \cite{a26} and references therein). 

At the moment, we pause for a while and give a brief proof that the
rainbow-ladder truncation scheme of DSE-BSE respects the scalar Ward
identity (Eq. (6)). Taking the derivative with respect to $m$ on both
sides of Eq. (15) and using again the identity Eq. (4), one arrives at
\begin{equation}
\frac{\partial G^{-1}(p,m)}{\partial m}=
\textbf{1}-\frac{4}{3}\int\frac{d^4q}{(2\pi)^4}g^2D_{\mu\nu}^{eff}(p-q)\gamma_{\mu}G(q,m)\frac{\partial
G^{-1}(q,m)}{\partial m} G(q,m)\gamma_{\nu}.
\end{equation}
Comparing Eq. (16) and Eq. (17), one sees that $\Gamma(p,0;m)$ and
$\partial G^{-1}(p,m)/\partial m$ satisfy the same integral equation
and therefore they must be equal, i.e., the scalar Ward
identity holds at the level of the rainbow-ladder truncation. That is to say, the
rainbow-ladder DSE-BSE model calculation respects the scalar Ward identity,
whatever form the effective model gluon propagator takes. In fact, the rainbow-ladder truncation scheme of
DSE-BSE itself is consistent in a broader sense.  In addition to
the scalar Ward identity, this truncation scheme preserves the
Abelian vector Ward-Takahashi identity and the axial-vector
Ward-Takahashi identity \cite{a23}. It is the preservation of the
latter identity that guarantees an understanding of chiral symmetry
and its dynamical breaking in the rainbow-ladder DSE-BSE framework
without a fine-tuning of model-dependent parameters \cite{a23}.

In this paper, we will
employ the rank-2 confining separable model gluon propagator
\begin{equation}
g^2D_{\mu\nu}^{eff}(p-q)\rightarrow\delta_{\mu\nu}D(p^2,q^2,p\cdot
q)=\delta_{\mu\nu}[D_0f_0(p^2)f_0(q^2)+D_1f_1(p^2)p\cdot qf_1(q^2)]
\end{equation}
with $f_i(p^2)=\exp(-p^2/\Lambda_i^2), i=1,2$. It is found to be
very successful in describing light flavor (including u, d, s-quarks)
pseudo-scalar and vector mesons with parameters $\Lambda_0=0.638~\mathrm{GeV}$, $\Lambda_1/\Lambda_0=1.21$, $D_0\Lambda_0^2=260.0$,
$D_1\Lambda_1^4=130.0$ and $m=0.0053~\mathrm{GeV}$ \cite{a27,a28}.

To proceed our model calculation, note that the quark propagator can
be decomposed in terms of two independent Lorentz structures (no
$\gamma_5$-related structure appears in pure strong interaction)
\begin{equation}
G^{-1}(p,m)=i\gamma\cdot pA(p^2,m)+B(p^2,m)
\end{equation}
Upon substituting Eqs. (18) and (19) into Eq. (15), the latter is converted
into two coupled integral equations ($s=p^2$)
\begin{eqnarray}
a(m)=\frac{D_1}{24\pi^2}\int^\infty_0ds\frac{s^2f_1(s)[1+a(m)f_1(s)]}{s[1+a(m)f_1(s)]^2+[m+b(m)f_0(s)]^2}
\\
b(m)=\frac{D_0}{3\pi^2}\int^\infty_0ds\frac{sf_0(s)[m+b(m)f_0(s)]}{s[1+a(m)f_1(s)]^2+[m+b(m)f_0(s)]^2}
\end{eqnarray}
with $A(p^2,m)=1+a(m)f_1(p^2)$ and $B(p^2,m)=m+b(m)f_0(p^2)$.

As for the dressed scalar vertex, its general form from Lorentz
structure analysis reads
\begin{equation}
\Gamma(p,0;m)=F(p^2,m)\cdot\textbf{1}+i\gamma\cdot pG(p^2,m),
\end{equation}
but only the first term provides the leading behavior \cite{a29}. We
shall keep only this term for simplicity. Putting this ansatz
$\Gamma(p,0;m)=F(p^2,m)\cdot\textbf{1}$, the model gluon propagator
Eq. (18) and the quark propagator Eq. (19) into Eq. (16), one has after
completing some algebras
\begin{equation}
\Gamma(p,0;m)=F(p^2,m)=1+\alpha(m)f_0(p^2),
\end{equation}
where $\alpha(m)$ satisfies the following equation
\begin{equation}
\alpha(m)=\frac{16D_0}{3}\int\frac{d^4q}{(2\pi)^4}f_0(q^2)\times[1+\alpha(m)f_0(q^2)]\times\frac{q^2A^2(q^2,m)-B^2(q^2,m)}{[q^2A^2(q^2,m)+B^2(q^2,m)]^2}.
\end{equation}
We point out that the Gaussian factors $f_0(p^2)$ and $f_1(p^2)$ in
the model gluon propagator provide sufficient ultraviolet
suppression and justify an effective cut-off for momentum
integration in Eqs. (20,21,24). Therefore no multiplicative
renormalization is needed.

With the dressed quark propagator ($A(p^2,m)$ and $B(p^2,m)$) and
then the dressed scalar vertex ($\alpha(m)$) solved, the
zero-temperature chiral susceptibility (Eq. (7)) can now be written as
\begin{equation}
\chi(m)=4N_cN_f\int\frac{d^4p}{(2\pi^4)}[1+\alpha(m)f_0(p^2)]\times\frac{p^2A^2(p^2,m)-B^2(p^2,m)}{[p^2A^2(p^2,m)+B^2(p^2,m)]^2}
\end{equation}
and calculated out. Of course, $\chi(m)$ takes an additive quadratic
divergence, which serves to cancel that one contained in the
temperature-dependent chiral susceptibility.

\subsection{The finite-temperature chiral susceptibility}
In recent years, the rainbow-ladder truncation DSE models are
extended to describe the finite temperature properties of QCD and
its chiral phase transition \cite{a25}. As for the foregoing
confining separable model, the extension to finite temperature is
systematically accomplished by transcription of the Euclidean quark
four-momentum via $q\rightarrow q_n=(\vec{q},\omega_n)$, where
$\omega_n=(2n+1)\pi T$ are the fermion Matsubara frequencies, and no
new parameters are introduced \cite{a28}. This means that the
effective gluon propagator at finite temperature is modeled as
\begin{equation}
g^2D_{\mu\nu}^{eff}(p_k-q_n)=\delta_{\mu\nu}[D_0f_0(p_k^2)f_0(q_n^2)+D_1f_1(p_k^2)p_k\cdot
q_n f_1(q_n^2)].
\end{equation}
with $p^2_k=\overrightarrow{p}^2+\omega^2_k$. The subsequent
calculations involving temperature-dependent dressed quark
propagator and scalar vertex will be parallel to those of the
zero-temperature case specified in the last subsection.

Firstly, the finite temperature version of the rainbow-DSE for the
dressed quark propagator Eq. (15) reads
\begin{equation}
G^{-1}(p_k,m)=i\gamma \cdot
p_k+\frac{4}{3}T\sum_{n=-\infty}^{+\infty}\int\frac{d^3p}{(2\pi)^3}g^2D_{\mu\nu}^{eff}(p_k-q_n)\gamma_{\mu}G(q_n,m)\gamma_{\nu}.
\end{equation}
Its solution now consists of three independent amplitudes due to the
breaking of $O(4)$ symmetry down to $O(3)$ symmetry \cite{a25}:
\begin{equation}
G^{-1}(p_n,m)=i\vec{\gamma}\cdot\vec{p}A(p_n^2,m)+i\gamma_4\omega_nC(p_n^2,m)+B(p_n^2,m).
\end{equation}
Substituting Eq. (28) into Eq. (27), one easily obtains solutions of
the form $A(p_n^2,m)=1+a(T,m)f_1(p_n^2)$,
$B(p_n^2,m)=m+b(T,m)f_0(p_n^2)$, and
$C(p_n^2,m)=1+c(T,m)f_1(p_n^2,m)$, where $a(T,m)$, $b(T,m)$ and
$c(T,m)$ are functions of $T$ satisfying the following three coupled
nonlinear equations:
\begin{eqnarray}
a(T,m)=\frac{8D_1}{9}T\sum_{n=-\infty}^{+\infty}\int\frac{d^3p}{(2\pi)^3}f_1(p_n^2)\vec{p}^2[1+a(T,m)f_1(p_n^2)]d^{-1}(p_n^2,m)
\\
c(T,m)=\frac{8D_1}{3}T\sum_{n=-\infty}^{+\infty}\int\frac{d^3p}{(2\pi)^3}f_1(p_n^2)\omega_n^2[1+c(T,m)f_1(p_n^2)]d^{-1}(p_n^2,m)
\\
b(T,m)=\frac{16D_0}{3}T\sum_{n=-\infty}^{+\infty}\int\frac{d^3p}{(2\pi)^3}f_0(p_n^2)[m+b(T,m)f_0(p^2_n)]d^{-1}(p_n^2,m),
\end{eqnarray}
where
$d(p_n^2,m)=\vec{p}^2A^2(p_n^2,m)+\omega_n^2C^2(p_n^2,m)+B^2(p_n^2,m)$.
These equations are numerically solved. Note that the confining
separable model quite facilitates summation over frequencies as well
as the three-dimensional integration owing to the sufficient
ultraviolet suppression guaranteed by the Gaussian functions
$f_0(p_n^2)$ and $f_1(p_n^2)$. As a result, no renormalization is
needed, which is the same as the zero-temperature case.

We now turn to the dressed scalar vertex. The finite temperature
version of the inhomogeneous ladder BSE for the dressed scalar
vertex (Eq. (16)) with vanishing total momentum reads
\begin{equation}
\Gamma(p_k,0;m)=\textbf{1}-\frac{4}{3}T\sum_{n=-\infty}^{+\infty}\int\frac{d^3p}{(2\pi)^3}g^2D_{\mu\nu}^{eff}(p_k-q_n)\gamma_{\mu}G(q_n,m)\Gamma(q_n,0;m)G(q_n,m)\gamma_{\nu}.
\end{equation}
The solution of Eq. (32) takes the general form
\begin{equation}
\Gamma(p_n,0,m)=F(p_n^2,m)\cdot\textbf{1}+i\vec{\gamma}\cdot\vec{p}G_l(p_n^2,m)+i\gamma_4\omega_n
G_t(p_n^2,m).
\end{equation}
The structure $F(p_n^2,m)$ makes the leading-order
contribution \cite{a30}. In order to be consistent with the
zero-temperature case and for simplicity as well, we shall keep only
this term and expect it would represent well the
temperature-dependent behavior of the dressed scalar vertex. Putting
this ansatz $\Gamma(p_n,0;m)=F(p_n^2,m)$, the model gluon propagator
(Eq. (26)) and the quark propagator (Eq. (28)) into Eq. (32), we
find
\begin{equation}
\Gamma(p_n,0;m)=F(p_n^2;m)=1+\alpha(T,m)f_0(p_n^2),
\end{equation}
where $\alpha(T,m)$ satisfies the following equation
\begin{equation}
\alpha(T,m)=\frac{16D_0}{3}T\sum_{n=-\infty}^{+\infty}\int\frac{d^3q}{(2\pi)^3}f_0(q_n^2)\times[1+\alpha(T,m)f_0(p_n^2)]
\times\frac{\vec{q}^2A^2(q_n^2,m)+\omega_n^2C^2(q_n^2,m)-B^2(q_n^2,m)}
{[\vec{q}^2A^2(q_n^2,m)+\omega_n^2C^2(q_n^2,m)+B^2(q_n^2,m)]^2}.
\end{equation}
Having obtained the amplitudes $A(q_n^2,m)$, $B(q_n^2,m)$ and
$C(q_n^2,m)$ from Eqs. (29-31)), we can numerically solve
$\alpha(T,m)$. The Gaussian function $f_0(p_n^2)$ appearing in the
numerator of the integrand justifies an effective cutoff for
summation over Matsubara frequencies as well as 3-D integration.

Now, with the dressed quark propagator and dressed scalar vertex at
finite temperature solved, the finite-temperature chiral
susceptibility Eq. (8) can be finally expressed as
\begin{equation}
\chi(T,m)=4N_cN_fT\sum_{n=-\infty}^{+\infty}\int\frac{d^3p}{(2\pi)^3}[1+\alpha(T,m)f_0(p^2_n)]\frac{\overrightarrow{p}^2A^2(p^2_n,m)+\omega^2_nC^2(p^2_n,m)-B^2(p^2_n,m)}{[\overrightarrow{p}^2A^2(p^2_n,m)+\omega^2_nC^2(p^2_n,m)+B^2(p^2_n,m)]^2}.
\end{equation}
The factor ``1'' (corresponding to the bare part of the dressed scalar
vertex) in the square bracket of the integrand would lead to
quadratic divergence, which will be removed by subtracting the
zero-temperature counterpart in our renormalization treatment. In
order for the finite-temperature chiral susceptibility to approach a
correct zero-temperature limit and then the temperature-independent
quadratic divergence to be canceled by the counterpart in the
zero-temperature chiral susceptibility, one should employ the same
numerical cut-off as that used in calculating $\chi(m)$ when
calculating $\chi(T,m)$, i.e. choose $(\overrightarrow{p}^2+\omega^2_n)_{max}=\Lambda^2_T\equiv
\Lambda^2_0$, where $\Lambda_T$ and $\Lambda_0$ denote the numerical
cut-off in calculating $\chi(T,m)$ (Eq.( 36)) and $\chi(m)$ (Eq. (25)),
respectively.

\subsection{Numerical results for renormalized chiral susceptibility}
Having obtained both the zero-temperature and finite-temperature
chiral susceptibility, we are now in a position to calculate the
renormalized chiral susceptibility. The numerical results are shown
in Fig. 1 and Fig. 2, where $\chi_R(T)$ are both scaled by $T^2$ and
hence dimensionless.

\begin{figure}[ht]
   \includegraphics[width=12cm]{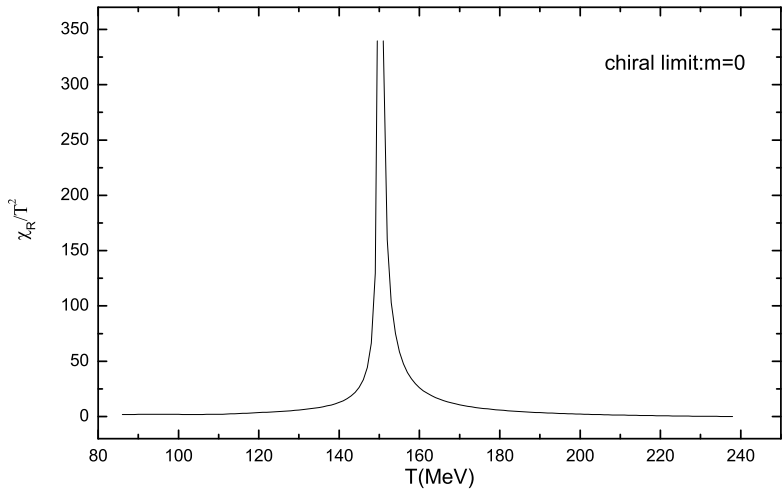}\\[-1cm]
  \caption{ $T$-dependence of the renormalized chiral susceptibility in the chiral limit.}\label{fig1}
\end{figure}

\begin{figure}[ht]
   \includegraphics[width=12cm]{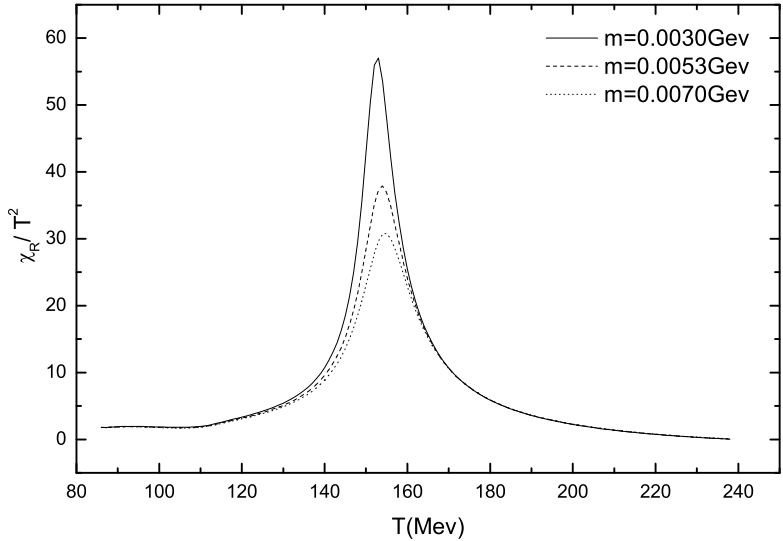}\\[-1cm]
  \caption{ $T$-dependence of the renormalized chiral susceptibility at different finite current quark masses.
  $m=0.0053~\mathrm{GeV}$ corresponds to the physical quark mass in our model.}\label{fig1}
\end{figure}

From Fig. 1, one sees that in the chiral limit, the renormalized
chiral susceptibility exhibits a very narrow, pronounced and in
fact, divergent peak at the chiral critical temperature
$T_c=150~\mathrm{MeV}$, which is a typical characteristic of phase transition
of second order driven by chiral symmetry restoration \cite{a10}.

Fig. 2 shows a quite different picture at finite current quark
masses ($m = 0.0053~\mathrm{GeV}$ corresponds to the physical value of the
degenerate u, d-quark current mass in our model \cite{a28}). There is
no singular behavior any more: the peak of the susceptibility
becomes not so sharp and pronounced as in the chiral limit; its
height is greatly suppressed and evidently finite; and there is no
unique temperature, but instead a range of finite width where the
transition phenomenon takes place. Also notable are the decrease in
the height of the susceptibility peak and that the susceptibility
peak has a minor shift from $150~\mathrm{MeV}$ to $155~\mathrm{MeV}$ as the current
quark mass increases. All of these observations agree well with
lattice QCD results \cite{a2,a6,a8,a12} and support an analytic
crossover involving a rapid change, as opposed to a jump, around the
pseudo-critical temperature.

\section{Conclusion and discussion}
In the present work, we present a continuum investigation of the
chiral susceptibility and try to say something about the character
of the two-flavor QCD thermal transition from the
temperature-dependent behavior of the susceptibility. We first
derive a model-independent integral formula, which expresses the
chiral susceptibility in terms of basic QFT objects: dressed
propagator and vertex. After appropriately regularizing the additive
quadratic divergence, we perform a model calculation of the
renormalized chiral susceptibility within the DSE-BSE framework with
a confining separable model gluon propagator which facilitates
summation over Matsubara frequencies. Our model study shows that in
the chiral limit, the renormalized chiral susceptibility exhibits a
narrow, divergent peak at chiral critical temperature $T_c=150~\mathrm{MeV}$,
signaling a second-order phase transition driven by chiral symmetry
restoration. On the other hand, at physical current quark masses,
the renormalized chiral susceptibility features a crossover, which
involves a rapid, but by no means singular, change in a temperature
range of non-vanishing width. So, encouragingly, the lattice QCD
observations \cite{a2,a6,a8,a12} are reproduced.

Compared to other continuum model studies, like DSE model
study \cite{a17} and NJL-type model study \cite{a15}, the present work
calculate the chiral susceptibility from an alternative
definition (the former two model studies defined the susceptibility
as the derivative of the generated mass with respect to the current
quark mass and hence the quadratic divergence was circumvented) and
in particular from a model-independent expression for the
susceptibility. Within the present framework for the chiral
susceptibility, the role of the dressed scalar vertex is
consistently included into account. However, the present
rainbow-ladder DSE-BSE model study still belongs to the class of
mean-field approximations, like the two model studies mentioned
above. It is interesting to investigate chiral susceptibility beyond mean-field approximations and this problem deserves further study.

\section{Acknowledgment}
We gratefully acknowledge the valuable communication with F. Karsch
on the quadratic divergence of chiral susceptibility and its
renormalization. Thanks also go to Lei Chang for enlightening
discussion on numerical calculations. This work was supported in
part by the National Natural Science Foundation of China (under
Grant Nos 10575050 and 10775069) and the Research Fund for the
Doctoral Program of Higher Education (under Grant No 20060284020).

\end{document}